# Treating Research Writing:
# Symptoms and Maladies

Varanya Chaubey

# Treating Research Writing: Symptoms and Maladies

Working in a field where revisions are inevitable, researchers often ask how they can improve a piece of writing.

There are two approaches to doing this: treat each problem that is visible on the surface or look deeper for the underlying malady that causes several problems to show up on the surface and treat that. Although either approach can be used, one may be more efficient for a particular case or at a particular stage in the writing process.

## 1. Treating Symptoms Visible on the Surface

Some problems in a piece of writing are visible on the surface. These are symptoms in the sense that we can identify them as deviations from the norm: the norm for the paper, for its category, or more generally, for good writing. For example, we can see that a paper is longer than others in its category or that a paragraph is more confusing than its neighbors. Such symptoms can be treated case-by-case.

This section contains six symptoms that readers often identify, even at a glance, along with some treatments for them. They are too-long sentences, fair sentences yet confusing paragraphs, too-long paragraphs, the lack of a visible storyline across paragraphs, a too-long paper, and too many footnotes.

### 1.1 Too-long sentences

Most writing guides recommend that English sentences have no more than 25 words. Longer sentences are harder to understand, so they may need to be distilled and, perhaps, restructured.

A too-long sentence, identified by a word count, can be distilled: many words can be replaced by a single word to convey the same meaning. Here are two examples:

| | |
|---|---|
| The telerium introduces protection, playing the role of a shield that separates the liminum from acids, and serves as an enclosure. | The telerium surrounds the liminum, protecting it from the acids outside. |
| 21 words | 11 words |
| In the event that a t-rail is forced out of its lodged position, it is to be expected that several o-rings will be allowed by it to fall. | If a t-rail is dislodged, it will likely drop several o-rings. |
| 28 words | 11 words |

A distilled sentence usually helps all readers form the same mental picture, efficiently. Take the example above. Here are the images that must be processed as we read the longer version of the sentence:

21 words

| The telerium introduces protection, | playing the role of a shield | that separates the liminum from acids, |
|---|---|---|
| 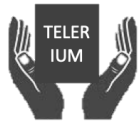 | 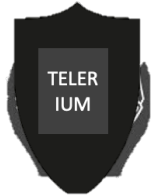 | 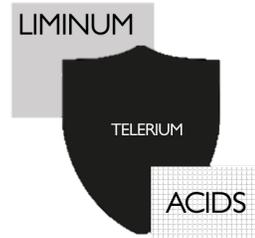 |

and serves as an enclosure.

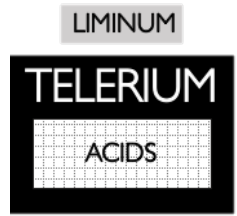 / 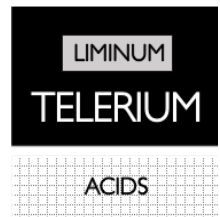

Here is the same idea in fewer words:

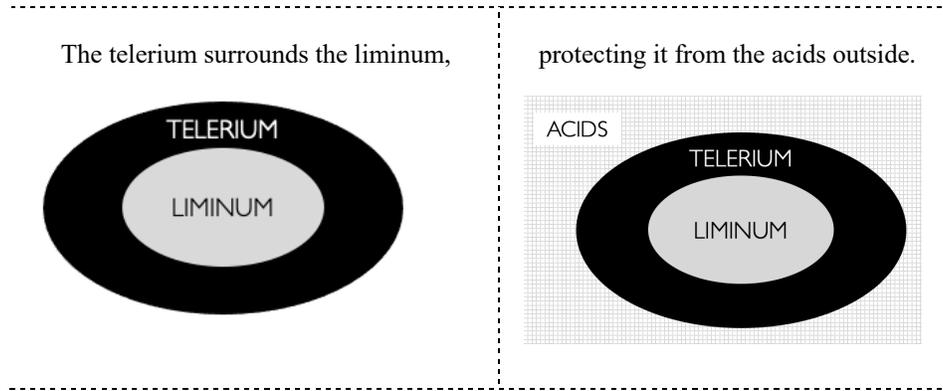

A single verb brought out of hiding can make a sentence shorter and more energetic. Verbs are the most direct way of telling the reader who did what. Yet there is an inclination among academic writers toward the preferential utilization of nouns with the resultant emergence of longer sentences.[1] You know you can bring a verb out of hiding when you see a long sentence with many nouns and a weak verb: some form of *to be*.

| The two topics of discussion were as follows: the consolidation of archives and the restriction of entry to members. | We discussed two topis: consolidating archives and restricting entry to members. |
|---|---|
| 19 words | 11 words |
| The increasing of the interest rate was carried out by the monetary authority. | The monetary authority increased the interest rate. |
| 13 words | 7 words |

You can treat longer sentences structurally too: look for the subject-verb-object core of the sentence. In English, it is recommended that the core be kept intact and close to the beginning of a sentence.

---

[1] As in this sentence. How might you revise it?

So, this is preferable:

> I read my book while sitting by the fire, even though it was a warm day.

to breaking up the core:

> I, while sitting by the fire even though it was a warm day, read my book.

or significantly delaying it:

> While sitting by the fire, even though it was a warm day, and listening to music, I read my book.

The exception is a dependent clause *briefly* delaying the core, such as:

> Even though it was a warm day, I read my book while sitting by the fire and listening to music.

As you restructure a long sentence, you may find you can break it up into a few shorter sentences. But which ideas should appear first? To work this out, think about the role the sentence must play in the whole: what it must do to advance the overall argument. The idea most relevant to that role goes into the first new sentence, and others can be put into later sentences. As you do this, you may also find that you get a chance to answer crucial logical questions that were going unanswered in the longer, poorly structured sentence.

| A lorimeter, which is cheap and easy to use but has been shown to deliver unreliable readings below 50 quartons and increasingly reliable readings above 100 quartons, recorded all readings in this study. | We chose a lorimeter to record our readings. It is cheap and easy to use. Moreover,…(whatever point the author was trying to make about the reliability of lorimeter readings in different ranges). |
|---|---|
| 33 words | 8, 7 and N words |

Relatedly, the underlying malady affecting a too-long sentence is often a missing or unclear link to the overall argument. If you start by articulating that link, the hierarchy of ideas and details usually becomes clearer: which details can be omitted, which ideas are secondary and can appear later, and which ideas must appear and appear early. For example, take this too-long sentence:

> We are the first to study the role of corotoid rupture over the life cycle using a Boyd-Barden style model with three t-points and data on both expected and realized stresses.

At 31 words, the sentence is longer than it should be. The word "first" hints that its role may be to help readers see the paper's contribution to the literature. But different readers interpret this differently:

1. Nobody else has studied the role of corotoid rupture over the life cycle, so they do it.
2. Nobody else has studied the role of corotoid rupture over the life cycle using a Boyd-Barden style model, so they do it.
3. To study the role of corotoid rupture over the life cycle, they adapt a Boyd-Barden model to have 3 (rather than $n \neq 3$) t-points.
4. They are the first to be able to study the role of corotoid rupture with data on *both* expected and realized stress variables.

As you see, each version relates to a slightly different overall argument. If the overall argument is kept in mind, it becomes easier to see what the sentence needs to say; then, it becomes easier to write a shorter, clearer sentence.

## 1.2 Fair sentences yet confusing paragraphs

What if our sentences are mostly clear, but the paragraphs they make up are not? This may be because our sentences are not ordered or linked properly.

As our elementary school teachers taught us, a good paragraph is, in one way, very much like a good sentence: it allows *all* its readers to form a clear picture—and indeed the *same* picture. We read "The girl kicked the ball," and we can all picture a female child striking, with her foot, a spherical object used for play.

The trouble is that even two individually clear sentences can yield a confusing picture when they are not properly linked. Take these two sentences: "The girl is sad. She saw her dog." Even if we all form the same picture from each sentence—a female child who is unhappy and a female child who at some prior moment saw her canine companion—we may form different pictures of the whole. Is the girl sad because she saw her dog, or is she sad even though she saw her dog? Five or six such sentences, taken together, can yield a very confusing idea of the whole.

So, to help join sentences into a clearer paragraph, we need to order and link them correctly.

***A. Order sentences dialogically***.  One way to order sentences is as an implicit dialog with readers.

Barbara Minto teaches us that a well-ordered piece of writing is not a monolog but an implicit dialog between the writer and her readers (Minto 1981). Something the writer says provokes a main question in readers' minds; this question is answered next, and, in turn, provokes a further main question; which is answered next and, in turn, provokes a further main question; and so on. Readers therefore participate as they read.  Even though they do not see their main questions articulated in the text, it is these questions that determine what appears next at every stage.

Without some dialog built in, readers can become confused about the meaning of the whole, even when they understand individual sentences. For example, consider this passage.

### Strosis

> If a t-rail is dislodged, it will likely drop several o-rings.  The speed coefficient is found to be between 1 and 2.  Under zero disturbance conditions, it is below 1.  Such strosis is observed in over 70 percent of cases.

Here, readers form reasonably clear images from each sentence they read, yet the whole is far from clear. To see why, let us study the logic of the order in which they appear. Specifically, as the reader makes her way through the text, what is the conversation she is having with the writer?

> WRITER: Strosis
> READER: OK, what's that?
> WRITER: If a t-rail is dislodged, it will likely drop several o-rings.
> READER: OK, is that strosis?
> WRITER: The speed coefficient is found to be between 1 and 2.
> READER: Is that strosis?
> WRITER:  Under zero disturbance conditions, it is below 1.
> READER: OK, but what is strosis?
> WRITER: Such strosis is observed in over 70 percent of cases.
> READER: Hello…is there intelligent life out there?

We can make this passage clearer by re-building it from scratch as a dialog with readers:

> WRITER: Strosis
> READER: OK, what's that?
> WRITER: Strosis is the phenomenon of elevated speed coefficients.
> READER: How does this happen?
> WRITER: In over 70 percent of cases, this happens when a t-rail gets dislodged and drops several o-rings on the thrusters. This causes thrusters to fire and elevates the speed coefficient to between 1 and 2.

The resulting passage is quite different from the original.  Which is easier to learn from?

| No Dialogic Links | Dialogic Links |
|---|---|
| <u>Strosis</u> | <u>Strosis</u> |
| If a t-rail is dislodged, it will likely drop several o-rings. The speed coefficient is found to be between 1 and 2. Under zero disturbance conditions, it is below 1. Such strosis is observed in over 70 percent of cases. | Strosis is the phenomenon of elevated speed coefficients. In over 70 percent of cases, this happens when a t-rail is dislodged and drops several o-rings on the thrusters. This causes thrusters to fire and elevates the speed coefficient to between 1 and 2. |
| 40 words | 43 words |

We have seen how the lack of dialogic links can make a little passage confusing to its readers—now, imagine how confusing a whole document structured as a monolog may be for readers. That is why, as Barbara Minto tells us, the whole document can and should be planned as a dialog with the reader. If your trouble is a paper that is not clear to its readers, you might consider going back to the drawing board and planning it that way.

***B. Include Explicit Links***. Beyond the implicit logical links reflected in the order of sentences, explicit words can be included to help readers join sentences into a single, meaningful whole.

To convey the logical relationship between two parts, we can include connector words and phrases. These include coordinating conjunctions (the so-called "fanboys": for, and, nor, but, or, yet, so); subordinating conjunctions (e.g., when, if, because, although, before); conjunctive adverbs (e.g., therefore, moreover, nevertheless).

The words we choose—both the connector word and the surrounding text—should leave no ambiguity about the logical link between the two parts being joined. For example, the connector word "and" can be ambiguous: it can be used to mean something that has happened either after or as a consequence of what has already happened or to introduce something of equal value to something that has already been mentioned; the context may not make clear which of these is meant. Similarly, "but" is used to contrast with what has already been mentioned, but the surrounding text may not make clear what the point of contrast is. Take this example:

> "Many studies have measured the effects of rotational adjustment cycles. But we are the first to study the dynamic effects of rotor adjustments on American machining and performance."

Here, even if readers understand that "but" is used to introduce a contrast, the surrounding text leaves some ambiguity about what the point of contrast is. Some readers assume it is dynamic versus static and others that it is geographic. For yet others, there are too many differences between the two sentences to see any single point of contrast. For example, is "study" the same as "measure" and is "the effects of rotational adjustment cycles" the same as "the effects of rotor adjustments"? If not, we cannot be sure of the point of contrast and therefore of the main contribution of this study.

To make the logical link — such as the point of contrast or connection — clear, you can deliberately repeat key terms from the previous sentence.

| *No repetition* | *Repetition for contrast* |
| --- | --- |
| Many studies have measured the effects of rotational adjustment cycles. But we are the first to study the dynamic effects of rotor adjustments on American machining and performance. | Many studies have measured the effects of rotational adjustment cycles on Chinese machining and performance. But we are the first to study the effects of these cycles on American machining and performance. |
| *No repetition* | *Repetition for connection* |
| Anori spores multiply rapidly in a process that causes perzocides to cluster and travel some distance. These attack the local fungi. | Anori spores multiply rapidly in a process that causes perzocides to cluster and travel some distance. These spores attack the local fungi. |

Deliberate repetition is especially helpful in linking larger chunks such as paragraphs and sections. For example, the repetition of a phrase across first sentences tells readers that an idea from a previous paragraph is being developed further; thus, that the story is advancing. Similarly, the repetition of a phrase across section takeaways tells readers that an idea from a previous section is being developed further.

## 1.3 Too-long paragraphs

In business writing, a regular paragraph should have about 5-6 sentences. Deviations from this norm can be identified and treated case-by-case. But keep in mind that several long paragraphs often mean there is a more serious underlying malady.

Regular paragraphs should be of similar length because they all have the same job: to advance the overall argument by one point. That point is stated in the first sentence; the other sentences, together, develop and support it. Business writers are advised to wrap up all of this in 5-6 sentences. One reason for this number may be that it is similar to the number of items the human working memory can hold. So, the idea may be that each point be supported by the number of details that the working memory can process (Miller 1956).

Finding a too-long regular paragraph is easy, and it also easy treat it once you have found it. At any point in the writing process, stop and scan a page. If you see a regular paragraph that looks bulkier than its neighbors, circle it. Such a paragraph signals that, *here,* there is something unusual is going on. Treatment can begin by identifying the point the too-long paragraph is trying to make in support of the overall argument. As Barbara Minto teaches us, this can be done by working from the bottom up and/or the top down:

*A. Treating from the bottom up:* Looking at all the details, articulate the point that these details, or a subset of them, can be said to support; check that this point belongs in the overall story.

For example, suppose you see these details:

1. Following political deadlock, the army took control for nearly five months between July 1994 and December 1994.
2. In 2000, as political instability continued, scheduled elections were postponed to 2001.
3. Between January 1995 and January 2000, there were three prime ministers and an interim government.
4. In 1999, political rallies were marred by violence and election violence claimed a total of 200 lives.

A point that these details could be said to support might be something like "The period between 1994 and 2001 was marked by political instability." If we go with this point, we know that the other details must be organized to support it. So, the sentence immediately following the first sentence could be the one talking about the political deadlock in 1994; then, the sentence about what happened in 1995, then the one about 1999, and finally the one about 2000. All that remains is to check that a point about political instability is relevant to the overall argument this paper makes.

The task of extracting a point from the details is harder when there are more details to sift through than the 4-5 expected in a single paragraph. Take these details:

1. A compromised strut can destabilize related elements in the machine and cause injuries.
2. Ivanovich says that dynamism can be improved by increasing pitch.
3. Ahmad advocates improving dynamism by using turbochargers.
4. Turbochargers have been proven to improve dynamism.
5. However, the cost of turbochargers is prohibitive for many in the industry.
6. Ivanovich's approach is inexpensive.
7. It is accomplished through an adjustment of the rear struts.
8. Even a small increase in pitch of .25 can lead to a 6 percent increase in dynamism.
9. However, Ivanovich's approach does not account for the greater load imposed by the on rear struts in the long term.
10. This greater load can compromise the structural integrity of the machine.
11. Our goal is to propose a cost-effective method to improve dynamism, without compromising the structural integrity of the machine.

In such cases, there may be more than one point being made:

¶ Existing proposals to improve dynamism are either prohibitively expensive or liable to compromise structural integrity.  [3]. [4] but [5]. [2+7].  [6] [9]. [10].

¶ [11]. [What type of supporting details could follow here?]

***B. Treating from the top down***: Working from the top down, articulate in one sentence the point readers expect to see next, given where they are in the overall argument. Construct the rest of the paragraph to support this point.

For example, say the point is, "But method X does not always yield reliable results." Knowing this gives us an idea of what details readers expect to see in the paragraph. For example, they may expect a sentence or two about the conditions in which method X is likely to fail; a sentence or two with examples of method X failing; and a sentence or two explaining why it fails in these conditions. We have 4-5 sentences in which to provide these details.

Whichever way you work—top down or bottom up—articulate the main point of the paragraph in a single sentence. Check that this point belongs in the overall story. Then, build the rest of the paragraph to support it.  Working this way will usually lead you to a more concise and more useful paragraph.

## 1.4 Lack of a visible storyline across paragraphs

In the workplace, readers often recognize the presence of a logical thread—or story—by a key term that is repeated across the first sentences of consecutive paragraphs.

For workplace readers, often the first indication that a story is being told is the appearance of a key term across consecutive first sentences.   Try it with the following excerpts: start by reading only the first sentence of every paragraph; then go back and read more carefully:

*EXCERPT 1:*
*Without Visible Storyline*

Consider situation 1.  In this situation, method X delivers efficient results.  It is….., which also allows it to be applied to situation 2…. Its….makes it useful in cases where…. Such features make it desirable in a variety of situations.  This includes situation 3, which is relevant for 80 percent of….

Method Y has some shortcomings.  It is… and ….  Moreover, when…., it can…But in one scenario it is more reliable than method X.  This scenario, Scenario 3, is detailed in Liu (2016).  Scenario 3 constitutes…. percent of all cases. The consequences of using method X in these cases has been…

Staff face a difficult decision when G is high and H is low.  So it is important to have a protocol for when to deploy which method.  We propose a new protocol for the combined use of methods X and Y.

*EXCERPT 2:*
*With Visible Storyline*

Nagarajan (2010) shows that method X has become the standard thanks to its convenient features.  It is….., which allows it to be applied to many…. It is also…., which makes it useful in cases where…. And it offers…., which is relevant for many…

But Liu (2016) presents a crucial scenario in which method X fails catastrophically while method Y delivers.  In Scenario 3, which constitutes … percent of all cases, X fails. The consequences of using method X in these cases has been…   In contrast, method Y, which is generally too cumbersome to use, is the only one that delivers results reliably. It….

We therefore propose a new protocol for the combined use of methods X and Y. This protocol is designed to help staff who have to make rapid decisions when G is high and H is low.

The repetition of a key term in consecutive first sentences tells readers that a story is being told and what it is. In Excerpt 1, there is no such term. In Excerpt 2, it is "method." Seeing that term in consecutive first sentences helps readers recognize that a single idea is being developed further and still further. It also helps readers, and even skimmers, understand something about the overall story, the one that these paragraphs, taken together, are meant to be telling. Even at a first glance, readers infer that that larger story is more likely to be about "method" than, say, "situation," "shortcomings," or "staff."

A preliminary check for the storyline—and any breaks in it—is therefore a quick scan of first sentences. Do successive first sentences have an explicit word or phrase in common? It does not have to be the same one all the way down the page, but often it is the same word or phrase linking several consecutive paragraphs before handing off to the next one; the handoff often occurs in a first sentence with both the old key phrase and the new one.

Fixing a break in the storyline can be easy or hard, depending on why the break exists. An easy fix is one where the paragraphs are, roughly, working: each one is making a point relevant to the overall story and the order in which these points are arranged is, roughly, logical. In such cases, the fix can be as simple as deliberately copying a phrase from one first sentence into another. There may be some work needed to adjust the other sentences in the affected paragraph, but if the fundamentals are good, this work is localized and quick. A harder fix is one where trying to create these links reveals that the section is not organized correctly; then, you may need to go back and plan paragraphs again from scratch.

## 1.5 A too-long paper

It is difficult to treat a too-long paper symptomatically without considering the underlying malady, except in one case: when you can lift out a chunk of text and find that the document still works.

An otherwise well-ordered paper can have one extraneous bit, a benign growth that, if you snipped it out, would not harm the rest. For example, perhaps there is a section that crept in at some point but, as the argument evolved, has become irrelevant. This could neatly be lifted out of the paper without disturbing the balance of neighboring sections. Or, there may be a few consecutive paragraphs that develop a tangential point and that you could lift out without compromising the main message of that section. In this case, all that may be needed is a little stitching and smoothing around the place the text was cut out.

But when a paper is said to be "too long," it could mean a few other things, and those things cannot be treated symptomatically. For example, it may mean that the paper is too long for what is normal at the journal or other outlet. Or it may mean that the paper is too long for what it says. In such cases, it is not efficient to treat excessive length by slashing bits here and there or by trying to distill the text sentence-by-sentence or paragraph-by-paragraph. It is better to address the underlying malady, for instance, to rethink the logic of the overall document, expressed in a few sentences, and build it up from there.

## 1.6 Too many footnotes

Too many footnotes, also, usually result from some underlying malady.

A footnote provides extra information that is not essential to the argument but likely to be of interest to a good chunk of the readership. In some disciplines, such as the humanities and law, the convention is to place all citations in footnotes.

What should *not* go into a footnote is any information relevant for assessing the argument being made. This is information which would, if it were not provided, render the reader's assessment of the argument incomplete or incorrect. One example is an analytical detail that the writer shrugs off as being minor but that most readers would say have a bearing on the argument. Another is a reference to the most relevant work in the literature, especially when that work has already made the bulk of the intellectual contribution being made in this paper.

What can go into a footnote is information that may be of interest to some readers. A footnote can pre-empt a tangential question many readers will have about a detail that is not essential to the argument at hand. It can be a reference to further reading that the writer thinks will be of interest to many readers, given what they are currently reading.

How many footnotes should there be? If they are not being used for all citations, as is the case in the humanities and law, a fair number may be about a third of the number of pages. So, in a 30-page paper, about 10, and in a 100-page document, about 33. What this means is that every three or so pages, you are inserting a piece of extra information for some part of the audience—this, I suspect is all they have the bandwidth to take in. If there are many more footnotes, ask why.

A document usually becomes overloaded with footnotes due to some underlying condition; address that, and many of the footnotes fall away. The most common condition is the writer's lack of confidence. This can stem from their life experience. Young researchers who have been subjected to illogical questions from posturing audience members begin worrying about pre-empting every question every reader could possibly have, including irrelevant ones. The lack of confidence can also come from not having fully understood the subject matter. The writer may not as yet be sure of what is essential and what is extra, but deadlines are forcing them to decide now. So, footnotes creep in.

An overuse of footnotes can also reflect the influence of outdated social norms on the writer. In some circles, an overabundance of footnotes seems to imply incredible knowledge. It conjures up the image of the genius, so filled with knowledge that he is unable to contain it. Freed from such notions, the writer may be able to see that many footnotes are unnecessary and can simply be deleted.

Unfortunately, footnotes can also creep in when unscrupulous gatekeepers to publication, such as referees, feel offended by new discoveries or pressure authors into citing irrelevant works.

# 2. Treating Underlying Maladies

We have seen how writers can work to treat individual symptoms. But if there are a handful of symptoms—the same problem in a few different places or a few different problems—it may be more efficient to look for and treat the underlying malady: a disorder that affects the whole system and gives rise to these symptoms.

An underlying malady may not be visible in the text, but it can give rise to a host of symptoms that, at first, seem unconnected. For instance, the malady of a weak overall argument can give rise to overlong sentences, confusing paragraphs, too many footnotes, and a too-long paper. Writers who stick to treating each of these symptoms individually are led a merry dance: they can spend hours, weeks, and even months treating each one separately without the draft getting much better.

This section describes common underlying maladies along with courses of treatment: a faulty overall argument, poor chunking, missing argument-relevance, mistaking rhetoric for logic, and warping pressures.

**Some Vocabulary.** Treating underlying maladies in writing is made difficult by the fact that the anatomy of a research paper is not well defined. We do not have well-established names for the structural elements in a paper, some of which may be unseen but are crucial to its health.

Other diagnosticians—doctors or plumbers, for example—have a shared professional vocabulary for such parts and what ails them— a blockage in the left anterior descending artery or in the main sewer line. This makes it easier to diagnose, discuss, and treat problems that are not visible on the surface. Without this, the task of bringing about improvements can be inexact, slow, emotional, and sometimes circular.

So, to aid the discussion here, I will borrow the names of parts I have used earlier (Chaubey 2017). Below is a brief description of each part along with its name.

| Name | Part |
|---|---|
| *RAP*: | The overall argument of a paper. It answers the reader's question of "And what is *your* contribution?" by positioning the paper in a space in the literature (*P*)—by saying what it is that is worth knowing that is as yet unknown—into which the research question (*R*) and answer (*A*) logically fit. When we set out to build a paper, we articulate each of these elements in a single sentence, but as we understand our paper better, we can often distill the *RAP* to still fewer sentences. |
| | The elements *R, A, and P* may be visible in the paper in their original form, but even if they are not, they are used to build it. For example, the 1-sentence *P* can be used to build out the first two paragraphs of the Introduction; the logic of those paragraphs is guided by the logic of the 1-sentence *P*. |
| | The *RAP* is a paper's identity: it is the essential logic of why a paper exists and what it does. It carries instructions for the development of sections, subsections, paragraphs, figures and tables in a paper. So, when it changes, the paper changes. |
| *RAP-relevant main point*: | The main idea yielded by any chunk of text in the paper — a section, subsection, or paragraph — in support of the overall argument. All the details in that chunk, when taken together, yield this idea that supports the *RAP*. |

## 2.1 A Faulty *RAP*

In my experience, the most significant malady in a research paper is a weak or unarticulated *RAP*. Associated symptoms include excessive length, poorly organized sections, and a pervading lack of clarity.

To understand why, let us back up to the first challenge in writing a paper: figuring out what idea the many details of project will coalesce around. Without such an idea, we simply have a collection of details rather than a coherent paper; readers cannot see the value of these particular details being brought together in this particular way. Even seasoned researchers take time to figure out what this idea must be for their paper.

One idea around which a research paper can be built is the *RAP*. This the logical combination of three elements— *R, A, and P* —that, together, answer the reader's question of "And what is *your* contribution?" *R* and *A*, the research question and answer, come straight from the research process. *P*, the positioning statement, shows why these are worthwhile, given the state of the literature: the body of useful knowledge we have accumulated so far. Each is articulated in just one sentence so that the writer is able to gain clarity about what her paper is. When a writer has articulated a logical *RAP*, she has found a way to organize the details, found a rule for determining what details go in and what details stay out.

Among faults within the *RAP*, perhaps the most problematic is a faulty *P*. *P* sets the trajectory for a paper. It defines the contours of what it is that is worth knowing that is as yet unknown, and by doing so, provokes *R*, the logical research question to pose in this context. *P* can fail for many reasons. For example, consider a *P* of the form "X has not been studied in the literature" which is meant to make space for a paper studying X. This *P* may fail if

- It does not represent the literature accurately: for instance, if X has been studied in some part of the literature.
- Readers do not know why the thing that is unknown is worth knowing. If X is the incidence of lung cancer, P works for many readers. We all know lung cancer kills people, so if the incidence of lung cancer has not yet been studied, it makes sense to study it. But if X is the possibility of the Woollcott-Zhu parameter being *strictly* greater than zero, many readers may not appreciate why this is worth studying.
- It sets the wrong expectations about what a paper delivers. Sometimes, a P of the form "X has not been studied in the literature" does represent the literature accurately and readers do know why X is worth knowing about. But what the paper delivers is an answer about X'. Then P may have provoked a research question and set expectations for some paper, but it is not *this* paper.

Sometimes, the problem is that the author is building a paper on two *P*s. They are trying to make two separate spaces in the literature and reel in two separate audiences by provoking two separate *R*s. This can lead to a two-headed introduction and a messy draft.

If you observe one or more of the following symptoms, it might be worth considering an underlying malady such as a faulty *RAP*: A paper does not seem to reach its audience. Readers find that they are unsure of what, exactly, the paper contributes to the literature or why it should have been written in the first place. The first few paragraphs in the Introduction do not get to the point. Readers find that the impression they formed of a paper's contribution early on is at odds with that they see later in the draft. Sections are disorganized. The draft balloons into something unwieldy as bits are added to satisfy different audiences. There are too many footnotes. Individual sentences are comprehensible, but it is unclear what ideas they yield when taken together.

A fault in the *RAP* can also lead to surprising symptoms, such as the overuse of adjectives and adverbs. A faulty *P* can manifest itself as the words "important" and "importantly" showing up 31 times in a 30-page draft. Because the writers are unsure of their overall argument, everything seems to them to be somehow — though it is unclear how — important. We can try to fix a symptom like that by using synonyms such as "significantly" and "interestingly," but the malady remains.

If you suspect that a faulty *RAP* is ailing your paper, it is usually efficient to put aside the draft and start again from scratch: by articulating a new *RAP* and building a matching outline. When you are pressed for time, this strategy may seem counter-intuitive; surely, it is easier to edit what is already on paper. I thought so too at first. But having watched the process unfold with over 800 researchers, I have been consistently informed that it is almost always slower and more confusing to tinker with something that was built on a faulty foundation than it is to start fresh.

## 2.2 Poor Chunking

In research writing, there is a reason we divide our text into chunks such as paragraphs and sections. A writer who is unsure of what that reason is, is likely to produce drafts that exhibit many symptoms of bad writing.

Chunking is the practice of grouping details so that they can be processed as a single idea.[2] Barbara Minto teaches us that we can think of the visible structural elements in a document —paragraphs, sections, and the document itself — as hierarchical chunks: many details grouped to convey a single idea and many ideas grouped to convey a single overall argument (Minto 1981).

We chunk to help readers process information, bypassing a limitation in the short-term memory. Unlike machines, human brains are not very good at absorbing individual details fired at them. If I were to recite just eight words at you, it is unlikely you would be able to repeat them back to me: cat, sponge, ladder, broccoli, koala, asparagus, juice, car.

George A. Miller explained this. The human working memory holds only seven items, plus or minus two (Miller 1956). To get around this limitation, we recode information into chunks —we group many details by some idea. This compression helps the brain keep track of more items than it otherwise could. The idea triggers recall of the many details housed in it.

Of course, if we are creating chunks to help our readers process information, *they* must see a meaningful association between the main idea of the chunk —which is usually visible at the top — and the details in it. So, Version B is a better way of chunking than Version A:

| Version A | Version B |
|---|---|
| <u>Happy</u> | <u>Dairy</u> |
| Eggplant | Milk |
| Milk | Yogurt |
| Cherries | Cream |
| | |
| <u>Rupa</u> | <u>Fruit</u> |
| Cream | Apples |
| Yogurt | Mangoes |
| Potatoes | Cherries |
| | |
| <u>Circumstantial</u> | <u>Vegetables</u> |
| Mangoes | Squash |
| Apples | Eggplant |
| Squash | Potatoes |

---

[2] For more on the uses of the word chunk, see Gobet, Lloyd-Kelly and Lane (2016).

The same goes for paragraphs. Version B is a better way of chunking than Version A:[3]

| *Version A* | *Version B* |
|---|---|
| G fluctuates between 5 and 7 percent. A 1- percent increase in J increases H by 6 per- cent. In the data, a 1-percent increase in J increases H by close to 12 percent. The cor- relation between F and K is 0.35. The cor- relation is expected to be above 0.7. In the data, G is seen to move between 2 and 10 percent. | The fit of the model is weak along three dimensions. First, it does not generate patterns of G observed in the data: it predicts that G fluctuates between 5 and 7 percent; in the data, G is seen to fluctuate between 2 and 10 percent. Second, H is less responsive to J in the model than in the data: a 1-percent increase in J only increases H by 6 percent, while in the data the number is closer to 12 percent. Third, the correlation between F and K, is low at 0.35; typically, it is expected to be above 0.7. |

If you approach a piece of writing by thinking about the principle for chunking, you can generally improve many chunks. Then, you know that there is a main idea visible at the top (such as in the first sentence of a paragraph or in the first paragraph of a section); and following this idea, details relevant to it (and no others).

Paragraphs and sections built this way are likely to yield a shorter, better-organized draft that others can learn from.

## 2.3 Missing RAP-relevance

Let us say we have good chunks in our paper. If these are not properly woven together, we are likely to see several symptoms of bad writing.

Ideally, each chunk in a paper will have some RAP-relevance that is visible to the reader. That is, readers will be able to see how the chunk is relevant to the overall argument: how the idea yielded by a particular grouping of details supports or develops the overall argument.

But this is not always the case. Take this brief paragraph at the top of the Data section in a paper. It is meant to show readers how the details in this section, taken together, support the argument.

---

[3] Example from (Chaubey 2017)

> In this section, we describe how we combine information from several sources to obtain an original dataset on the earnings of all public-school teachers in the period 1974-1994. This data includes a group of more than 200,000 individuals that can be easily identified as private tutors, since their salaries are sourced from a private tutoring institution. We link this dataset with information on classroom outcomes to provide a comprehensive picture of the extent to which incentives shape teaching in the public sector, since it identifies teachers' conflicting incentives in a clear way and it contains information on all teachers in the public sector, including those who have a second job as a private tutor.
>
> <div align="right">114 words</div>

In class, PhD candidates and advanced researchers spend up to 30 minutes trying to draft a 1-2-sentence answer to the question of how, exactly, this section supports the overall argument; not all arrive at it. The adjectives hint at it — words like "clear" and "comprehensive" —but are too vague to be helpful. In general, weak or missing RAP-relevance wastes readers' time and is a missed opportunity for the writer.

If writers bear in mind that the RAP-relevance must be explicit, they may instead write a paragraph like this:

> We want to study whether teachers in public schools face conflicting incentives when they are permitted to freelance as private tutors. So, we link two datasets: one with the earnings of public-school teachers, where we can identify income from private tutoring institutions; and the other with classroom outcomes for these teachers.
>
> <div align="right">51 words</div>

This may not be perfect, but it is an improvement. Readers can more easily see why the dataset is apt, given the research question.

Clarifying the RAP-relevance of a chunk often helps writers see that organizing the details differently may yield an even stronger argument. For example, the writer may realize that if the details in Sections 3.3 and 4.2 were combined into a single chunk, the RAP-relevance of the resulting chunk would be greater. Such reorganization leads to a shorter, sharper paper.

## 2.4 Mistaking rhetoric for logical argument

Working your way up in the social sciences, you may have heard people speak of rhetoric as equivalent to good writing. Others are sceptical and with good reason: rhetoric can be persuasive without being logical. If we are prone to rhetoric, we can offer ourselves a few logical checks as we write.

What rhetoric *is*, for most of us, is somewhat nebulous. In Plato's *Gorgias*, a dialog about the nature of rhetoric, Socrates refers to it as an experience in producing delight and gratification, a part of flattery, and a means to persuade the ignorant. In Aristotle's *Rhetoric*, it is defined as the faculty of observing in any given case the available means of persuasion. It is also described as the counterpart of dialectic as well as an offshoot: rhetoric uses non-argumentative means of persuasion, which are not used in dialectic. For most of us, this is not easy to follow. The one-line dictionary definition is "the art of speaking or writing effectively." But what does that mean, exactly?

What rhetoric *does* is better defined. Its effect, included in most dictionary definitions, is that the reader or listener is persuaded: moved to believe that something is true. This says nothing about whether the thing they are persuaded to believe is true or untrue, logical or illogical. In *Gorgias*, Plato brings up this point.

> **Socrates.** … the rhetorician need not know the truth about things; he has only to discover some way of persuading the ignorant that he has more knowledge than those who know?
>
> **Gorgias.** Yes, Socrates, and is not this a great comfort?

Engaging in rhetoric is at odds with the collective goal of research writing: to record, share, and advance knowledge. But the skillful use of rhetoric can certainly be mistaken for good research writing, even by the writer herself. Good research writing is clear and complete. Rhetoric may be pleasing and persuasive, even when it is unclear or incomplete. It may not be the most straightforward way of saying something. More words may be used than necessary, making it harder for readers to trace the logical skeleton that lies beneath.

Luckily, rhetoric has tell-tale signs, one of which is the lack of concise answers to logical questions. For example, in the early parts of a document, the reader may have a logical question about the meaning of a term—an implicit request for a concise definition—and not receive it. In *Gorgias*, Socrates and Chaerephon set out to define what, exactly, a rhetorician like Gorgias is: what the nature of his art—rhetoric—is. They receive an answer that is rhetorical in that it sounds great but does not actually answer the question:

> **Chaerephon.** But…what is the art in which he is skilled?
>
> **Polus.** O Chaerephon, there are many arts among mankind which are experimental, and have their origin in experience, for experience makes the days of men to proceed according to art, and inexperience according to chance, and different persons in different ways are proficient in different arts, and the best persons in the best arts. And our friend Gorgias is one of the best, and the art in which he is a proficient is the noblest.
>
> **Socrates.** Polus has been taught how to make a capital speech, Gorgias; but…he has not exactly answered the question which he was asked…
>
> **Polus.** Why, did I not say that it was the noblest of arts?

> **Socrates.** Yes, indeed, but that was no answer to the question…. nobody asked what was the quality, but what was the nature of the art.

As pleasing as it is, writing that employs rhetoric does not pass logical scrutiny. For example, consider the interim report written by an ad hoc committee formed in 2017 to consider a code of professional conduct for the American Economics Association. I am including excerpts here, but you can and should read the whole document for yourself. It is available here: https://www.aeaweb.org/resources/member-docs/code-of-conduct-interim-report.

> In particular, the AEA should aspire to a professional environment that promotes equal opportunity…. Moreover, the Committee believes that the AEA mission can only be fulfilled if economists adhere to the highest level of honesty and integrity in all aspects of their work.
>
> Recent events raise concerns about systematic deviations from these ideals…*[Terms pulled from the paragraphs that follow: unacceptable behavior, a culture that is hostile towards women, some economists engage in bullying, bigotry, sexism, racism, or other forms of discriminatory behavior towards their colleagues or students, a hostile or confrontational climate in the workplace, including at seminars and conferences, anecdotal evidence of economists who abused their professional status, abused processes related to promotion, abused their personal or professional relationships, failed to disclose personal or professional relationships, or failed to disclose relevant personal financial interests].*
> .
> These practices indicate breakdowns of the honesty and integrity that should guide economists' professional conduct.
> .
> .
> .
> **Recommendation**:
> The committee recommends that the American Economic Association adopt a Code of Professional Conduct. Draft language for this code is attached for consideration by the AEA Executive Committee.

At this point, readers have logical follow-up questions:

> What is the professional code of conduct you recommend? And how did you come to choose it?

They expect to see logical answers — at least overview answers — to these questions in the text that follows; later they will look at the draft language in more detail. Here is the text that follows:

> The committee reviewed a number of codes of conduct implemented by other associations in economics and social science including those of the American Finance Association, the European Economic Association, and associations for other social sciences. These codes reflect thoughtful consideration of many of the issues that confront economists in their professional conduct. The Committee ultimately favored a more parsimonious statement of principles for several reasons:

Now, readers' logical question grows more urgent: "Well *why*, after reviewing codes you describe as reflecting "thoughtful consideration of many of the issues that confront economists in their professional conduct" did you *ultimately* favor a more parsimonious code of conduct?"

And here is the text that follows:

> - The parsimonious statement of mission in the AEA by-laws has served the association well for more than a century, and serves well as rubrics for the key principles of professional conduct that the committee recommends

But if this were true, why would this committee have been appointed, this very report have been called for in the first place? How could this be true and the statement that appears earlier in the text *also* be true — that there have been "systematic deviations" from the ideals? Or that "unacceptable behavior has been allowed to continue through tacit toleration," practices that "indicate breakdowns of the honesty and integrity that should guide economists' professional conduct," including people who "abused their professional status, abused processes related to promotion, abused their personal or professional relationships, failed to disclose personal or professional relationships, or failed to disclose relevant personal financial interests."

Why, you might wonder, would anyone place such a statement *here* just where the reader expects a logical answer to a fundamental question?

The power and the danger of rhetoric is that it arouses emotions, moves us to believe what we are reading or hearing, even without fully understanding it. Large numbers, by their sheer weight, are especially good at achieving this. In this statement above, there is a reference to "more than a century." If anything, that statistic undercuts the point being made: a serious problem has developed in those very 100 years and perhaps as a result of the outdated nature of the very thing being lauded. But the weight of 100 years, placed on us by eminent scholars in the field, is not easy to shrug off. Laboring against it is the gentle suggestion from the back of our minds that there may be a logical gap in what we are reading or hearing; and it does not always win.

So, what can we do about this? For writers who believe that rhetoric does not have a place in research writing, there are simple things we can do. To begin with, we can check when we have provoked a logical question in readers' minds.

Then, we can work to offer a concise answer: try to formulate it one sentence spoken out loud. Conciseness is no guarantee against rhetoric. But it does help us (and others) examine the underlying logic, to spot gaps in reasoning. For example, if we had sat down to tea with the committee that ultimately favored the parsimonious code of conduct and asked them to tell us in just one sentence why they did so, they might have replied simply: "Because we weren't given the time and resources to do better." In the document itself, here is a sentence that appears a whole 235 words after the logical question was raised, a long sentence buried at the bottom of the section, wearing the look of an afterthought:

> As a practical matter, if the AEA decides it wants to adopt a detailed professional code of conduct along the lines of other associations, it would need to appoint a committee with sufficient time to prepare such a document, including provisions for collecting suggestions and feedback from the profession.

As a mark of respect for our readers, let us make it a habit to articulate concise answers to logical questions and to deliver these answers promptly; that is, at the moment when readers need them to assess the logic of what we are saying. Details can follow, fleshing out this concise answer. But as with any chunk, let us make sure the leading text is a concise version of the answer, the essential idea expressed in a few words. For instance, we can try articulating 1-sentence answers to the following questions and place them where readers expect to find them in our documents:

- What is….? (any element of the work, such as model/data/setting/empirical strategy)
- Why is this a good ….? (what its qualities are)

## 2.5 Pressure

A paper is what it is. Still, writers frequently feel compelled to dress it up to look like something else. This rarely results in good writing.

Discovery is a complex process, to the extent that researchers don't always know what a paper *is* until years after it is written and published. Sure, they may have an idea and write it up following that idea. But when some years have gone by, they may look back and realize that, actually, it could have been written quite differently. We all get to know our work differently once time has gone by.

But that process can be warped by pressure: someone or something compelling us to write the paper not following our most natural sense of what it is but in a way that makes it seem like something else. For example, I regularly hear sentences that go as follows: "My main contribution is A, but my advisor feels that for the job market, I should really make it about B" or "Even though my paper is not directly related to X, I need to make people care, so I am saying a lot about it" or "These parts don't naturally fit together, but I was told I need to show off all my skills on the market, so I am putting everything in the draft." These are examples of pressure warping a paper out of its logical form.

We learn how to bend to such pressure in our student days, even when doing so is neither in our own interest, nor in the interests of our fields. If you have worked with graduate students, you have probably had the unfortunate privilege of watching this happen for the first time. Under pressure, especially from advisors, researchers say things that their better judgment in other times would probably prevent, from the ridiculous — "Well, my findings support scholar X's theory, but my advisors said we don't like scholar X, so I am trying to write the Introduction in a way that makes it non-obvious" — to the worrying — "Problem Y with my analysis undercuts my whole argument, so I am burying it on page 14 not to make it too obvious."

Strategic pressure warps research writing more than we would want it to, a problem that is not helped by the current system of gatekeeping in some fields, including in economics. Unfortunately, there is no convenient demographic we can point to and say, "Maybe that is a problem *there*, but it would never happen *here* at my department or my journal". Nor can we be sure that once a researcher learns to bend to warping pressures, that they will learn to stand firm in the future.

Our collective writing will improve when those who decide things for our field—tenured faculty, journal editors, and hiring and promotion committees — find ways to alleviate these pressures. (And when they set up committees to do so, it is only fair to expect that the written communications these committees make public do not suffer from the same pressures they are setting out to alleviate.)

Of course, one can hope that competitive forces in science will ensure that it is only logical writing that rises to the top — but this is not guaranteed. For example, if enough people who see no problem with rhetoric in scientific writing happen to be at the top, through the forces of homophily highlighted in Akerlof and Michaillat (2018), they may help others with the same outlook rise to the top. This is why we need a conscious effort to educate and remind each other of what good scientific writing is, what is logical and what is not. We cannot just hope that good writing will, automatically and always, prevail.

In the meantime, we can take steps of our own. We can lay bare the logic of our own drafts for others to assess. We can commit to describing things exactly as they are. And we can commit to actually answering the questions that others ask.